\documentclass[prl,aps,twocolumn,showpacs]{revtex4}

\usepackage{graphicx}
\usepackage{bm}

\def\ket#1{|#1\rangle}
\def\bra#1{\langle #1 |}
\def\e{\varepsilon}
\def\expect#1{\langle #1 \rangle}
\def\w{\omega}
\begin{document}

\title{Can competition between crystal field and Kondo effect cause
non-Fermi liquid like behavior?}

\smallskip

\author{Frithjof B.~Anders$^1$  and Thomas Pruschke$^2$}
\affiliation{$^1$
  Theoretical Physics, University of the Saarland, 66041
  Saarbr\"ucken, Germany\\ 
             $^2$Institute for Theoretical Physics,
             University of G\"ottingen, D-37077 G\"ottingen, Germany}

\date{\today}


\begin{abstract}
The recently  reported unusual behavior of the static and dynamical
magnetic susceptibility as well as the specific heat in
Ce$_{1-x}$La$_{x}$Ni$_{9}$Ge$_4$ has raised the question of a possible
non-Fermi liquid ground state in this material. We argue that for a
consistent physical picture the crystal field
splitting of two low lying magnetic doublets of the  Ce $4f$-shell
must be taken into account. Furthermore, we show that for a splitting
of the order of the low temperature scale $T^*$ of the system 
a crossover behavior between an SU(4) and an SU(2) Kondo
effect is found. The  screening of the two doublets occurs on different
temperature scales leading to a different behavior of the magnetic
susceptibility and the specific heat at low temperatures. The
experimentally accessible temperature regime down to 50mK still lies
in the extended crossover regime into a strong coupling Fermi-liquid
fixed point. 
\end{abstract}

\maketitle


\paragraph{Introduction.}
The investigation of thermodynamic and transport properties of strongly
correlated electron systems is of fundamental importance for our
understanding of elementary excitations in solid state
physics. Especially measurements on the metallic heavy fermion (HF) compounds
\cite{Grewe91} challenge the paradigm of  Landau's Fermi-liquid
concept which incorporates all lattice and Coulomb correlations into a
renormalized quasi-particle mass $m^*$ and a few Fermi-liquid
parameters.

In many cases the presence of localized moments in the HF
compounds leads to magnetic or superconducting phase transitions,
which either compete with each other in Ce based compounds or possibly
even coexist as in Uranium based materials. 
The experimental evidence \cite{Maple95,Stewart01} compiled over the
past ten years also indicates  that even for HF systems  with
paramagnetic ground state, the temperature dependence of the specific
heat and the magnetic susceptibilities often does not agree with the
predictions of the Fermi-liquid theory, in particular when subject to
pressure or ion substitution \cite{Loehneysen96}. Therefore, the
phenomenological term non-Fermi liquid (nFL) was attributed to such
regimes appearing in a large variety of different materials
\cite{Maple95,Stewart01}.

The understanding of the observed nFL  behavior is one of the most
challenging and unsolved theoretical puzzles. In many materials, it
is ascribed to a quantum critical point (QCP) at which a transition
temperature is suppressed to $T=0$ by an external control parameter
such as pressure or doping\cite{Loehneysen96,Steglich96}. It is
believed that in the vicinity of such a QCP, quantum fluctuations
dominate over thermal ones even at finite temperatures as shown by
Hertz\cite{Hertz76}  and Millis\cite{Millis93} in a renormalized
quasi-particle picture. Despite a tremendous experimental and
theoretical effort it is, however, still not clear whether the nFL
effects observed in HF compounds are related to novel low-lying
non-local excitations in concentrated systems, true local nFL physics
or simply due to competing local energy scales. 


Recently, the experiments showing unusual specific heat, magnetic
susceptibility, and resistivity data for  Ce$_{1-x}$La$_{x}$Ni$_{9}$Ge$_4$
for various concentrations have drawn a lot of attention
since this material has the ``largest ever recorded
value of the electronic specific heat at low temperature''\cite{CeNi9Ge4} 
of $\gamma(T)= \Delta C/T \approx 5 {\rm JK}^{-2} {\rm
 mol}^{-1}$. While the $\gamma$ coefficient continuous to rise at the
lowest  experimentally accessible temperature, the magnetic
susceptibility apparently tends to saturate at low temperatures.
Experimentally, the quantum critical\cite{Millis93}  and Kondo
disorder scenario\cite{KondoDisorder1995} were ruled out \cite{CeNi9Ge4}.

In this Letter, we propose a local scenario for the observed nFL
behavior in  Ce$_{1-x}$La$_{x}$Ni$_{9}$Ge$_4$. This is backed by the
experimental findings that the electronic contribution to the specific
heat as well as the magnetic susceptibility normalized to the Ce
concentration remains almost independent of the La
concentration\cite{CeNi9Ge4}. We will show that the competition of
Kondo and crystal field effects leads to a crossover regime connecting
incoherent spin scattering at high temperatures and a conventional
strong-coupling Fermi-liquid regime at temperatures much lower than
the experimentally accessible 30mK.

\paragraph{Crystal-field scheme for CeNi$_{9}$Ge$_4$.}
The Hund's rule ground state of Ce$^{3+}$ with
$j=5/2$ is split in a tetragonal symmetry\cite{CeNi9Ge4} in three
Kramers doublets. If the  crystal electric field (CEF) parameters are
close to those of cubic symmetry, the two low lying doublets
$\Gamma_7^{(1)}$ and $\Gamma_7^{(2)}$, originating from the splitting
of the low  lying $\Gamma_8$ quartet, are well separated from the higher lying
$\Gamma_6$ doublet. Ignoring this $\Gamma_6$ doublet, we can discuss
two extreme limits. In a cubic environment, the  CEF splitting vanishes
and the low temperature physics is
determined by an SU(4) Anderson model described by a strong coupling
fixed point plus a marginal operator responsible for the particle-hole
asymmetry \cite{KrishWilWilson80b}. In a strongly tetragonally
distorted crystal, on the other hand, the crystal field splitting of
the quartets is expected to be large. In this case, the low
temperature properties are determined by an SU(2) Anderson model which
has a {\bf significantly lower Kondo scale} since the degeneracy $N$ enters
the denominator of the exponential $T_K\propto \exp[-1/N J]$. Then the
second doublet at higher energies is screened at temperature $T\approx \Delta =
E_{\Gamma_7^{(2)}} -E_{\Gamma_7^{(1)}}$ and contributes little to the
magnetic susceptibility. Thus, the experimental 
response would be that of a simple SU(2) Anderson model which was 
ruled out by the experiments \cite{CeNi9Ge4}. Therefore, we propose
that the material parameters lie in the crossover regime where the
effective low temperature scale $T^*$  is of the order of the crystal field
splitting $\Delta$. Then,  the excited doublet will have significant
weight in the ground state so that the total magnetic response
differs  from a simple SU(N) Anderson model.

\paragraph{Formulation.}
Our calculation is based on an SU(N) Anderson model with infinite-$U$ 
\cite{Toulouse1969} whose Hamiltonian is given by
\begin{eqnarray}
  \label{eq:siam-cef}
H&=&  \sum_{k\alpha} \e_{k\alpha\sigma} 
c^\dagger_{k\alpha\sigma} c_{k\alpha\sigma} 
+ \sum_{\alpha\sigma} E_{\alpha\sigma}
\ket{\alpha\sigma}\bra{\alpha\sigma} 
\\
&& +
\sum_{k\alpha\sigma} V_{\alpha\sigma}\left( 
\ket{\alpha\sigma}\bra{0}  c_{k\alpha\sigma} 
+c_{k\alpha\sigma}^\dagger \ket{0}\bra{\alpha\sigma}
\right)
\nonumber
\; ,
\end{eqnarray}
where $\ket{\alpha\sigma}$ represents a state with energy
$E_{\alpha\sigma}$ on the Ce $4f$-shell of the $\alpha$-th irreducible
representation (irrep)  with spin $\sigma$ and $ c_{k\alpha\sigma}$
annihilates a conduction electron state with energy
$\e_{k\alpha\sigma}$ transforming according to the irrep
 $\alpha$ of the tetragonal magnetic point
group\cite{Cox87}. This allows locally only fluctuations between an
empty and a singly occupied Ce $4f$-shell.

We accurately solve the Hamiltonian (\ref{eq:siam-cef}) using Wilson's
numerical renormalization group \cite{Wilson75,KrishWilWilson80a} best
suited to deal with the competition between Kondo effect and CEF field
splittings. The key ingredient in the NRG is a logarithmic
discretization of the continuous bath, controlled by the parameter
$\Lambda > 1$~\cite{Wilson75}. The Hamiltonian is mapped onto a
semi-infinite chain, where the $N$th link represents an exponentially
decreasing energy scale $D_N \sim \Lambda^{-N/2}$. Using this
hierarchy of scales the sequence of finite-size Hamiltonians ${\cal
  H}_N$ for the $N$-site chain is solved iteratively, truncating the
high-energy states at each step to maintain a manageable number of
states. The reduced basis set of ${\cal H}_N$ thus obtained is
expected to faithfully describe the spectrum of the full Hamiltonian
on the scale of $D_N$, corresponding to a temperature $T_N \sim D_N$
\cite{Wilson75} from which all thermodynamic expectation values are
calculated. 

In order to obtain the impurity contribution to the specific heat, we
calculate the difference between the entropy of the full model
(\ref{eq:siam-cef}) and the corresponding free electron gas
$S_{free}(T)$,
i.e.\ $\Delta S(T) = S_{tot}(T) -S_{free}(T)$ \cite{Wilson75}. 
The Sommerfeld coefficient $\gamma(T) = \Delta C(T)/T$ of the
Ce contribution to the specific heat is directly obtained by
differentiating $\Delta S(T)$ with respect to $T$:
\begin{eqnarray}
  \Delta C(T) &=& T \frac{\partial S(T)}{\partial T}
\, 
\end{eqnarray}
Since $\Delta C(T)/k_B$ is a dimensionless quantity, the NRG
predicts the absolute magnitude of $\Delta C(T)$ per Ce without any
further fitting parameter. The experiment, however, is needed to
provide the absolute scale of the temperature axis. Since 
$\gamma$ has the dimensions of an inverse energy, we obtain an
estimate of the 
Anderson width $\Gamma_0=V^2\pi \rho(0)$ by  comparison with
the experiment\cite{ScheidtKehreinPruschkeAnders2005} ($\rho(0)$ is
the density of states of the conduction electrons at the chemical
potential). While the calculation allows for different values of the
hybridization matrix element $V_\alpha$ to incorporate different
coupling strength of the multiplets to the conduction band, we used
the same matrix element $V$ for all calculations to keep the number of
free parameter as small as possible.

Assuming a Zeeman splitting of the multiplet energy
$E_{\alpha\sigma}$ as $E_{\alpha\sigma} = E_{\alpha} -
\sigma g_\alpha \mu_B H$, the Ce contribution to the magnet susceptibility is
given by
\begin{eqnarray}
  \Delta \chi &=& \mu_B^2 \sum_\alpha g_\alpha^2 \frac{\partial
    \expect{S_{\alpha}^z} }{\partial \bar H_\alpha}  
= \mu_B^2 \sum_\alpha g_\alpha^2 \chi_{\alpha}
\end{eqnarray}
where the magnetic field $\bar H_\alpha$ is the
Zeeman splitting energy $\bar H_\alpha = g_\alpha \mu_B H$. While
the $g$-factor is determined by the CEF states of the multiplets, we
view them as adjustable parameters and calculate  $\chi_{\alpha}$ by applying
an very small external magnetic field of the order of $\bar H/\Gamma_0
=10^{-9}$ and estimate the values of
$g_\alpha$  by comparing with the experiment
\cite{ScheidtKehreinPruschkeAnders2005}. As a consequence, the dynamical susceptibility
$\chi''(\w)$ presented later does not contain any adjustable parameter and the $q$
integrated dynamical structure factor $S(\w)$
 can be obtained by the dissipation fluctuation theorem,
\begin{eqnarray}
S(\w)\left[1-e^{-\beta\w}\right]  &=& \chi''(\w)
=  \mu_B^2 \sum_\alpha g_\alpha^2 \chi_{\alpha}(\w)
\;\; ,
\end{eqnarray}
where $\chi''_\alpha(\w)$ is given by the imaginary part of the spin
susceptibility 
$  \chi_\alpha (\w)  = \ll S_{\alpha}^z |  S_{\alpha}^z \gg (\w)
$.
Since the two doublets contribute differently to $S(\w)$ for
temperatures comparable or lower than the CEF splitting $\Delta$, the
proposed fit by a simple  Lorentzian for $\chi''(\w)$ must obviously
fail \cite{SwCeNiGeSces05}. At temperatures well above the splitting, the
magnetic response of the two doublets is this of a quartet, and in
that regime a single Lorentzian fit is justified. At $T<\Delta$,
$S(\w)$ cannot be approximated by a Lorentzian consistent with the
reported neutron scattering data  \cite{SwCeNiGeSces05}.



 \begin{figure}[tb]
 \centering
 \includegraphics[width=85mm,clip=true]{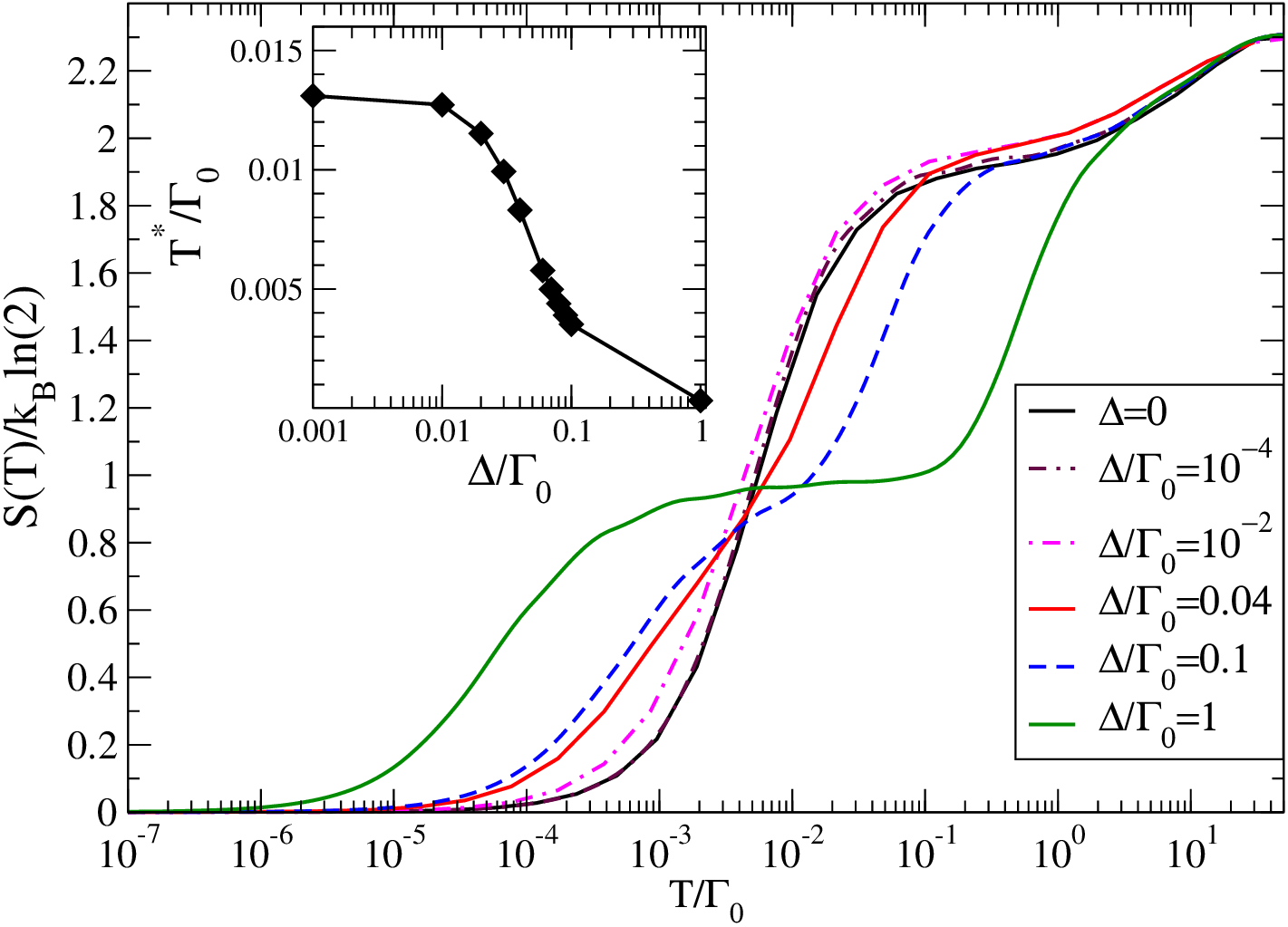}
 \vspace{-5pt}
 \caption{The impurity contribution to the entropy in units
   $k_B\ln(2)$  for a ground state doublet
   $E_{\Gamma_7^{(1)}}=-8\Gamma_0$ and a band    width of $50\Gamma_0$
   as function of the CEF splitting $\Delta$. The inset shows the
   splitting dependent low temperature scale $T^*(\Delta)$. The
   calculations were performed with NRG discretization parameter
   $\Lambda=4$, keeping $N_s=1500$ states at each NRG step.
   }
\label{fig:entropy}
 \end{figure}

\paragraph{Results.}

The impurity contribution to the entropy
is plotted as function of temperature for various CEF splittings
$\Delta$ in Fig.\ \ref{fig:entropy}. A symmetric band of  $D=50\Gamma_0$ with a constant density
of states $\rho_0=1/2D$ is used in all calculations. While the entropy
curves are very similar for small  splitting, for larger $\Delta/\Gamma_0>
0.05$ the decrease of the entropy occurs in two steps from
$\ln(4)\to\ln(2)$ and from $\ln(2)\to 0$ when decreasing the
temperature. This implies that the screening of the two doublets occurs at
two separate energy scales once the splitting $\Delta$ exceeds the low
temperature scale $T^*(\Delta)$. We have defined the scale
$T^*(\Delta)$ \cite{Wilson75} as the temperature at which the impurity
contribution to the effective spin moment $\Delta \expect{S^2_z}
=\expect{S_z^2}_{tot}-\expect{S_z^2}_{free}$ is reduced by a factor of
2 compared to its local moment value of $1/4$. The inset shows the
splitting dependency of this low temperature scale. Once the splitting
reaches a value comparable to $T^*(0)$, the low temperature scale is
rapidly reduced. For splittings slightly large than
$T^*(0)/\Gamma_0=0.0131$, the crossover region of the entropy from
$\ln(4)\to 0$ is significantly extended compared to the SU(4) curve as
shown by the solid red (online) curve for 
$\Delta/\Gamma_0=0.04$. The onset occurs at higher temperature due to earlier
screening of the upper doublet while the strong coupling Fermi-liquid
fixed point is reached at much lower temperatures compared to a simple
Kondo model. In Ce$_{1-x}$La$_{x}$Ni$_{9}$Ge$_4$ this temperature
regime was phenomenologically attributed to a nFL behavior
\cite{CeNi9Ge4}. Based on our calculations, we, however, argue that
a rather extended crossover regime  to a  Fermi-liquid fixed point is
observed in the experiments.

We achieved the best agreement between theory and experiment in the
Kondo regime of (\ref{eq:siam-cef}) using a ground state doublet
energy of  $E_{\Gamma_7^{(1)}}/\Gamma_0=-8.5$ and a splitting of
$\Delta/\Gamma_0=0.015$. We have used the comparison of the
dimensionless experimental and theoretical specific heat to obtain the
absolute scaling factor for the temperature axis, and the
$\gamma(T)$-coefficient to assign an explicit value to
$\Gamma_0$. Both procedures gave $\Gamma_0\approx 714K=61.6meV$ and,
therefore, a bare CEF splitting of $\Delta=10.7K$ was used in the
calculations. The corresponding entropy curve would be located between
$\Delta/\Gamma_0=0.01$ and $\Delta/\Gamma_0=0.04$ in
Fig.~\ref{fig:entropy}. No lattice renormalization effects have been
taken into account since  the experiments indicate very good scaling with
the Ce concentration \cite{CeNi9Ge4}. The additional Schottky peak
observed at $T\approx 60K$ in the experimental data stems from the
third doublet neglected in our calculation.

\begin{figure}[tb]
 \centering
 \includegraphics[width=89mm,clip=true]
 {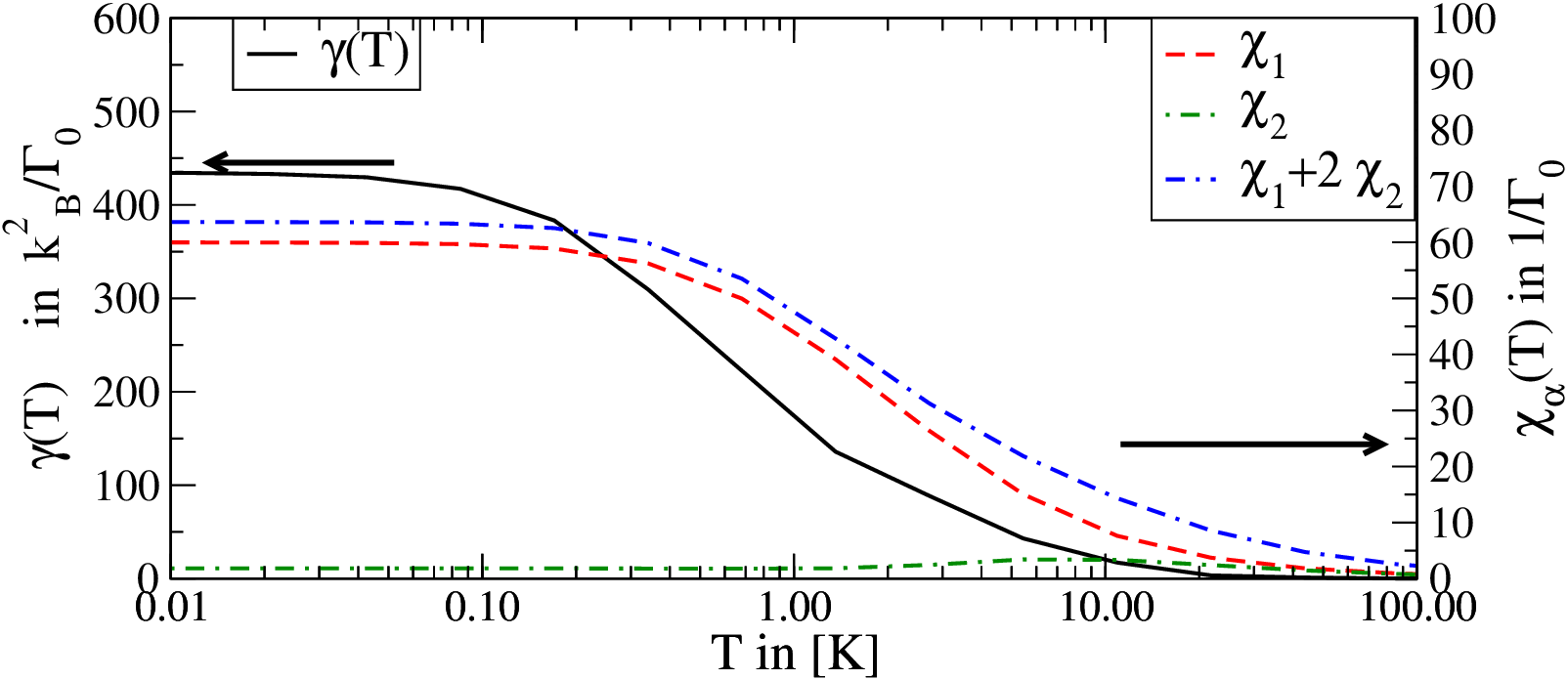}
 \vspace{-5pt}
 \caption{Comparison between $\gamma(T)=C(T)/T$  vs $T$ and the
   susceptibility contributions of the two doublet vs $T$ for
   $E_{\Gamma_7^{(1)}}/\Gamma_0=-8.5$, $\Delta/\Gamma_0=0.015$. The 
   contribution of the lower doublet, $\chi_1$ is much larger than the
   one of the upper doublet, $\chi_2$. NRG parameters as in Fig.\ 
   \ref{fig:entropy}.
   }
   \label{fig:dc-sus}
\end{figure}

The comparison between the temperature dependence of $\gamma(T)$ and
$\chi(T)$ is shown  in Fig.\ \ref{fig:dc-sus} assuming a ratio
of $g_2^2/g_1^2=2$ for a good fit to the experimental
data \cite{ScheidtKehreinPruschkeAnders2005}. The ground state doublet
dominates the magnetic response at low temperature and tends to
saturate at temperatures higher than the $\gamma$-coefficient,
consistent with the experiments \cite{CeNi9Ge4}. We 
find this behavior only for CEF-splittings $\Delta\approx T^*(\Delta)$
while for much larger or much smaller values $\chi(T)$ and $\gamma(T)$
saturate simultaneously.

\begin{figure}[tb]
 \centering
 \includegraphics[width=89mm,clip=true]
 {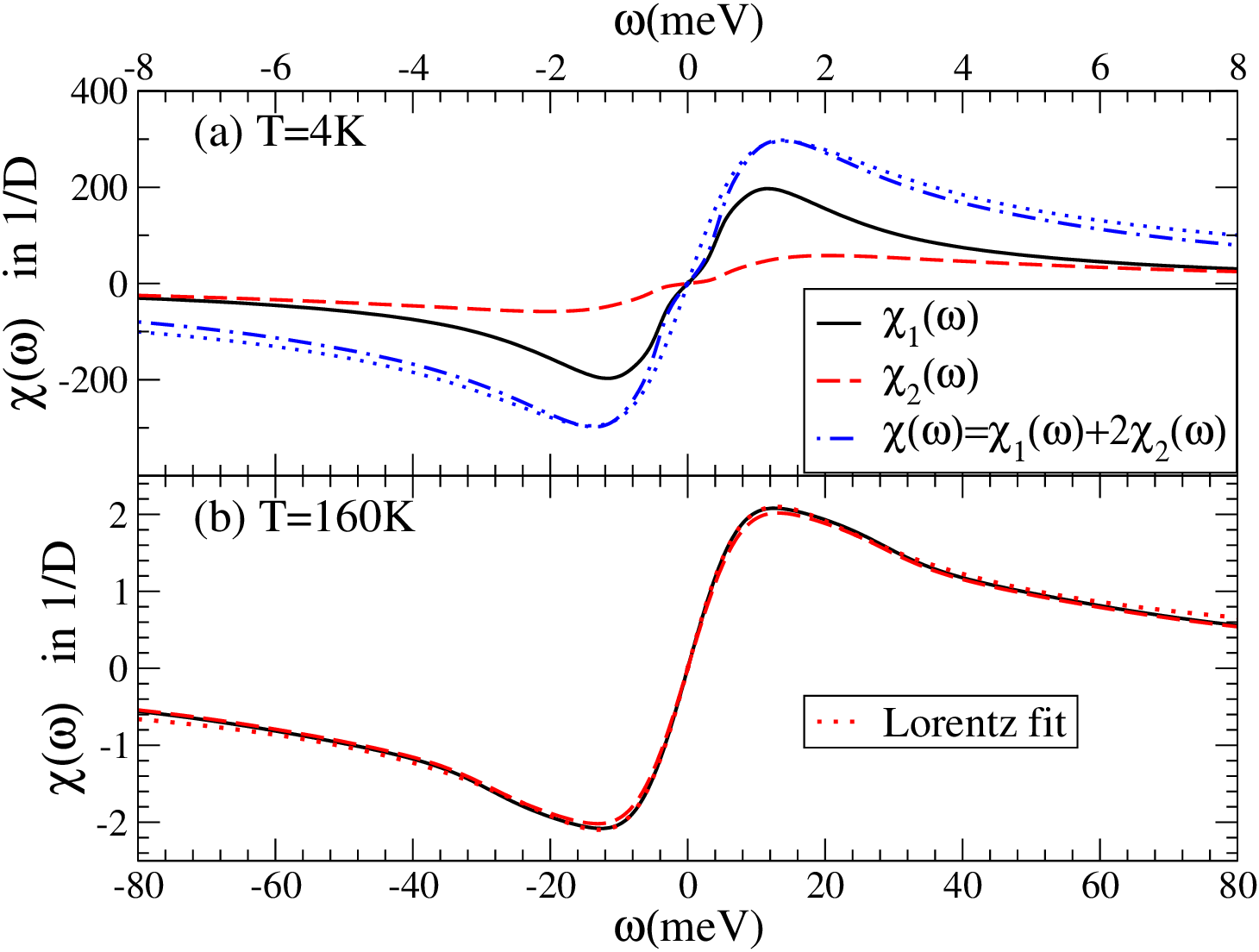}
 \vspace{-5pt}
 \caption{Comparison of the contributions to $\chi''(\w)$ at (a) $T=4K$
   and (b) $T=160K$ for the model parameters of Fig.\
   \ref{fig:dc-sus}. The dotted lines present a Lorentzian
   fit. Calculations were done with NRG discretization 
   $\Lambda=2.5$ keeping $N_s=6400$ states at each step.
   }
   \label{fig:chi-w}
\end{figure}

Having obtained  a reasonable estimate for  the ratio between the $g$
values of the doublets, we can predict the temperature dependence of the
imaginary part of the local  magnetic susceptibility $\chi(\w)$. Our
results are shown in Fig.\ \ref{fig:chi-w} for (a) $T=4K$ and (b) at
$T=160K$ for the parameter set of 
Fig.\ \ref{fig:dc-sus}. While for high temperatures (panel b) the
contributions to $\chi''(\w)$ are identical for both doublets and can
be fitted by a Lorentzian $\chi''(\w) \approx A_0 \frac{\w\Gamma(T)
}{\w^2 +\Gamma^2(T)}$ of width $\Gamma(T)=13meV$, this is  not
possible for $0.1K<T<40K$ as seen in (a). Only at temperatures below $0.1K$,
$\chi''_1(\w)$ might be described by a Lorentzian using a $\Gamma$ of
$1meV$. An effective line width of $\Gamma=1.2meV$ was used to obtain
the dotted (online blue) line in (a) fitting $\chi''(\w)$. In contrast
to a naive ionic picture, the ground state contains significant
contributions from the first excited doublet.

\begin{figure}[t]
 \centering
 \includegraphics[width=89mm,clip=true]
 {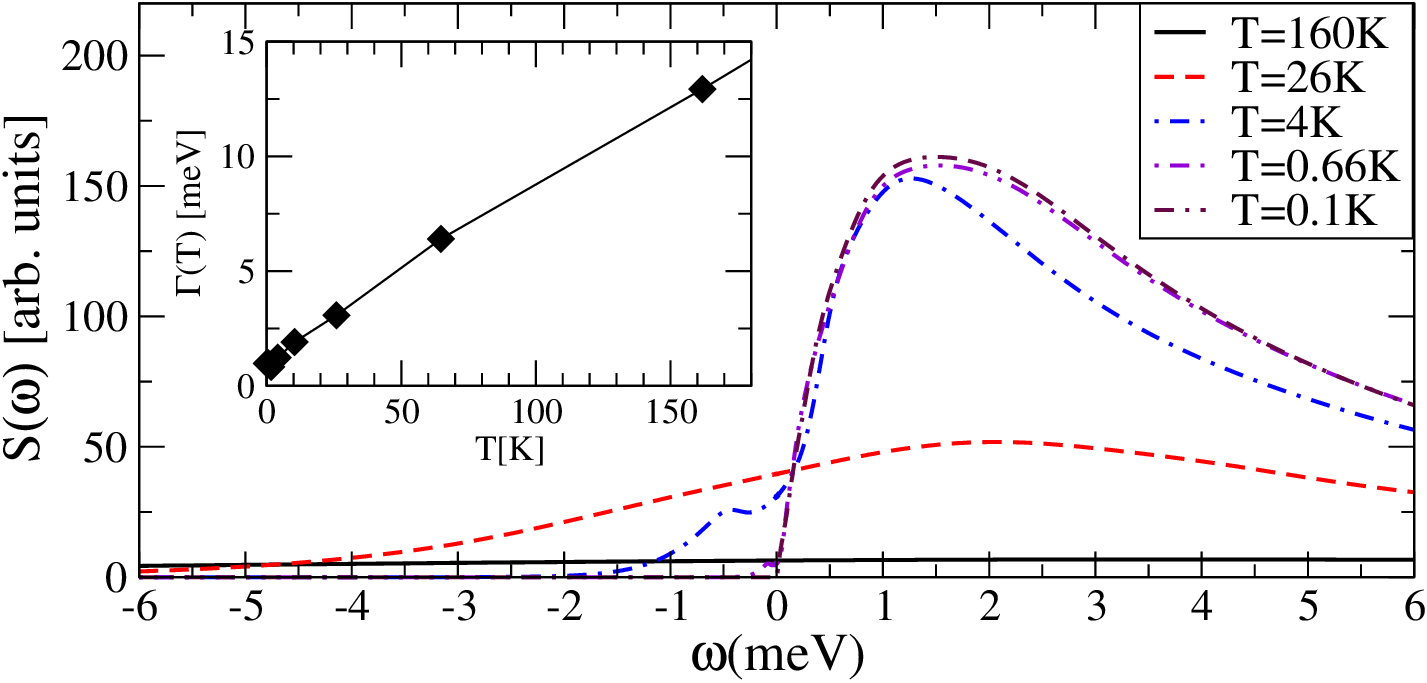}
 \vspace{-5pt}
 \caption{The $q$-integrated dynamical structure factor $S(\w)$ for
   different temperatures versus frequency for NRG parameters as in
   Fig. \ref{fig:chi-w}. The inset shows an estimate for temperature
   dependent relaxation rate $\Gamma(T)$ fitted to the lower doublet only.
   }
   \label{fig:sofw}
\end{figure}

Our findings naturally explain the failure of the attempt to fit
recently reported neutron scattering data on power samples of
concentrated CeNi$_9$Ge$_4$ with a simple Lorentzian for temperatures
below $30K$ \cite{SwCeNiGeSces05}. The change of slope in $\chi''(\w)$
yields a  small peak for $\w<0$ in $S(\w,T=4K)$ shown in Fig.\
\ref{fig:sofw}. Note, that the unusual behavior of $\chi_\alpha(\w)$
in the crossover regime can also not be explained as originating from a distribution of Lorentzians \cite{Bernhoeft2001} used to fit $S(\w)$ in
\cite{SwCeNiGeSces05}. Such phenomenological approaches contain
obviously limited amount of information if no consistent physical
picture for {\em all} physical properties emerges. A rough estimate of
$\Gamma(T)$ plotted as inset to Fig.\ \ref{fig:sofw} indicates
$\Gamma\propto T$ as in the experiment \cite{SwCeNiGeSces05} before it
saturates below 100mK. Our absolute values for $\Gamma(T)$
are roughly a factor of two larger than those reported in
\cite{SwCeNiGeSces05}. This is consistent with the fact that $T^*$ for
Ce$_{0.5}$La$_{0.5}$Ni$_9$Ge$_4$, the material we used to fix our
model parameters, is larger that $T^*$ for  CeNi$_9$Ge$_4$.
In addition, our error in the estimate of the CEF splitting $\Delta$ sensitivly
determines the low temperatur scale $T^*$ as depicted in the inset of
Fig. \ref{fig:entropy} and therefore our absolute energy scale.

\paragraph{Summary and discussion.}
A simple physical picture of the competition between Kondo effect and
CEF splitting leads to an extended crossover region from the high
temperature free multiplet to the low temperature strong coupling
fixed point if the crystal field splitting is of the order of the low
temperature scale $T^*$. Both doublets contribute significantly
to the magnetic response at the fixed point yielding {\em different}
contributions to the static and dynamic susceptibility. We propose
that the origin of the nFL behavior in Ce$_{1-x}$La$_{x}$Ni$_9$Ge$_4$
is related to this extended crossover region compared to a simple
Kondo model or a two channel Kondo lattice scenario
\cite{AndersJarCox97,CoxZawa98}. This provides a consistant picture
for the temperature dependence of specific heat and the magnetic response
in agreement with {\em all} experimental data in all regimes. We hope
to include lattice coherence effect in future calculations to explain
the transport properties as well.

We have benefited from discussions with E.\ W.\ Scheidt.
F.B.A. acknowledges funding of the NIC, Forschungszentrum J\"ulich,
under project no. HHB000, T.P.\ financial support by the Deutsche
Forschungsgemeinschaft through Sonderforschungsbereich 602 and
supercomputer support by the
Gesellschaft f\"ur wissenschaftliche Datenverarbeitung G\"ottingen and
the Norddeutsche 
Verbund f\"ur Hoch- und H\"ochstleistungsrechnen.


\end{document}